\documentclass[preprint,prc,showpacs,preprintnumbers,amsmath,amssymb,floatfix]{revtex4}
\usepackage{amsmath}

\usepackage{graphicx}
\usepackage{dcolumn}
\usepackage{bm}
\usepackage{ulem} 
\usepackage[usenames]{color}
\usepackage{epstopdf}
\usepackage{epsfig}
\usepackage{float}
\usepackage{subfigure}
\usepackage{multirow}
\usepackage[version=3]{mhchem}

\newcommand{\nc}{\newcommand}       
\nc{\vc}[1] {\mbox{\boldmath $#1$}} 
\nc{\del}       {\partial}              
\nc{\bra}       {\langle}               
\nc{\ket}       {\rangle}               
\nc{\bras}[1]   {\langle #1|}           
\nc{\kets}[1]   {|#1\rangle}            
\nc{\mapleft}[1]{           
 \smash{\mathop{\,          %
  \hbox to 1.5cm{\rightarrowfill}\, }\limits_{#1}}}
\nc{\beq}     {\begin{eqnarray}} \nc{\eeq}    {\end{eqnarray}}
\nc{\nn}      {\\\nonumber} \nc{\vs}      {\vspace{-0.275cm}}
\nc{\fra}    {\frac{1}{2}}
\nc{\mb}        {\mathbf}

\begin{document}

\preprint{}

\title{Neutron stars within relativistic central variational method}

\author{Jinniu Hu}
\email{hujinniu@nankai.edu.cn}
\affiliation{School of Physics, Nankai University, Tianjin 300071,  China}
\author{Hong Shen}
\affiliation{School of Physics, Nankai University, Tianjin 300071,  China}
\author{Hiroshi Toki}
\affiliation{Research Center for Nuclear Physics (RCNP), Osaka University, Ibaraki, Osaka 567-0047, Japan}

\date{\today}
\begin{abstract}
The properties of neutron stars are investigated within the relativistic central variational method by using a realistic nucleon-nucleon ($NN$) interaction. The strong repulsion of realistic $NN$ interactions at short distances is treated by a Jastrow central correlation function, whose form is completely determined through minimization of the total energy of the nuclear many-body system. The relativistic Hartree-Fock wave functions are chosen as the trial wave function. In this framework, the equation of state of the neutron star matter in $\beta$ equilibrium is obtained self-consistently. We further determine the properties of neutron stars via the Tolman-Oppenheimer-Volkoff equation using Bonn A, B, and C potentials. The maximum masses of neutron stars with these realistic potentials are around $2.18 M_\odot$ and their corresponding radii are around $11$ km. These results are in accordance with the calculations of the relativistic Brueckner-Hartree-Fock theory with the same potentials. Furthermore, we also find that the splitting of proton-neutron effective masses will be reversed at high density in the neutron star matter, which are caused by the contribution of short-range correlation on kinetic energy.
\end{abstract}



\maketitle

\section{Introduction}
The first neutron star was discovered by Bell and Hewish in 1967. Since then, the neutron star has spawned the topics of research in numerous studies in astronomy and nuclear physics. A neutron star can be treated as a huge nucleus composed of neutrons, protons, and leptons~\cite{glendenning97,weber99}. The properties of neutron stars, like mass and radii, can be precisely measured with the gradual development of astronomical observation technology. This development has promoted theoretical studies on the nuclear many-body theory~\cite{prakash97,heiselberg00}, especially on the properties of nuclear matter at high densities, where high density refers to densities above 3 times the nuclear saturation density in the core of a neutron star.

In the Tolman-Oppenheimer-Volkoff (TOV) equation~\cite{oppenheimer39,tolman39}, which is derived by solving the Einstein equations for a spherically symmetric and time invariant system, the properties of a neutron star are fixed by the equation of state (EOS) of neutron star matter, which is a charge neutral system in $\beta$ equilibrium. The EOSs of neutron star matter have been widely studied using various nuclear many-body theories, by the application of density functional theory (DFT) and {\it ab initio} methods in nuclear physics~\cite{dutra12,dutra14}.

The effective nucleon-nucleon ($NN$) interactions are adopted in DFT, which are determined by fitting the empirical saturation properties of nuclear matter and the ground states of finite nuclei. There are usually two schemes used in DFT, i.e., relativistic and non-relativistic approximation. The most popular nuclear many-body theory in non-relativistic DFT is the Skyrme-Hartree-Fock theory~\cite{vautherin72,bender03,stone07}, which was constructed based on the point-coupling $NN$ interaction. In the relativistic version, covariant DFT was proposed based on the one-boson-exchange picture~\cite{serot86,ring96,meng06}. The EOSs of DFT around the saturation density region can be constrained very well and are kept consistent within different effective $NN$ interactions. However, their behaviors in the high density region, especially in the cores of neutron stars, show variability, which generate widely different predictions of the properties of neutron stars. With the discoveries of massive neutron stars, whose masses are around $2M_\odot$, many DFT interactions are eliminated~\cite{demorest10,antoniadis13}.

On the other hand, in the state-of-the-art nuclear many-body methods, via {\it{ab initio}} calculations, a realistic $NN$ interaction is used that is obtained by reproducing the $NN$ scattering data. The most popular realistic $NN$ potentials are constructed by the meson-exchange picture~\cite{machleidt89} and chiral effective field theory~\cite{epelbaum09,machleidt11}. The $NN$ potentials, based on chiral low-momentum expansions, are soft cores at short distances due to pion exchange; whereas, in the meson-exchange potential, such as the Bonn potential, the $\omega$ and $\rho$ meson exchanges are essential in the core region of the $NN$ interaction, which generate very strong repulsive core. There are also other high-momentum effects at short distance, such as the strong tensor force, besides strong repulsion~\cite{jastrow55} which is related with the nucleon inner structure~\cite{toki80,oka81} in the realistic $NN$ interaction. Since 1950, many microscopic many-body methods have been proposed to take such high momentum contributions into account. Akmal {\it et al.} used the hyper-netted chain-summation technique in the non-relativistic variational method~\cite{pandharipande79} and the AV18 potential~\cite{wiringa95} to calculate the EOS of neutron rich matter~\cite{akmal98} with a three-body force. Krastev and Sammarruca adopted the relativistic Brueckner-Hartree-Fock (RBHF)~\cite{brockmann90} theory to study the properties of neutron stars with Bonn potentials~\cite{krastev06}. Gandolfi {\it et al.} also compared their results with the calculations made by Akmal using the auxiliary field diffusion Monte Carlo method~\cite{gandolfi09}. Hebeler {\it et al.} discussed the properties of neutron star with $NN$ interactions from chiral effective filed theory and constrained the neutron star mass-radius relation in the framework of renormalization group theory~\cite{hebeler10,hebeler13}. All of these values of EOSs obtained from {\it{ab initio}} calculations agree well with each other in the high density region and generate similar properties of massive neutron stars.

For neutron rich matter, the tensor effect in realistic $NN$ interaction is suppressed largely in the $T=1$ isospin channel \cite{krastev06,hu10,wang12,hu13}. Therefore, it is enough to explicitly consider only the short-range correlation in neutron rich matter. Based on this motivation, a relativistic central variational (RCV) method was proposed by including a central correlation Jastrow function~\cite{hu11}, which was inspired by the series works by Panda {\it{et al.}}~\cite{panda05,panda06,panda07}. In the RCV method, the solutions of the relativistic Hartree-Fock (RHF) method~\cite{bouyssy87} were chosen as the trial wave function of the total system. The saturation properties obtained by the RCV method are $20\%$ different from the empirical data, likely because of the lack of tensor correlation in symmetric nuclear matter. However, the EOS of pure neutron matter within the Bonn potentials can reproduce the results from RBHF theory very well. The solutions of the RHF method in nuclear matter are constructed by spinors with plane wave functions, the summation of spin and isospin for which can be easily and systematically evaluated via the Feynman trace technology. Therefore, the calculation in the RCV method for neutron-rich matter is essential and economical, when compared with other time-consuming {\it{ab initio}} methods such as non-relativistic variational methods and the RBHF theory in asymmetric nuclear matter.

Furthermore, the calculation procedures of the {\it{ab initio}} aforementioned methods, when they were applied to neutron star matter in $\beta$ equilibrium, are complicated, therefore, the equations of $\beta$ equilibrium are usually solved with various approximations instead of self-consistence treatments~\cite{akmal98,krastev06}. Hence, the intention of this work is to apply the RVC method to calculate the EOSs of neutron star matter in $\beta$ equilibrium self-consistently and use these EOSs to study the properties of neutron stars with realistic $NN$ interaction, Bonn potentials. In addition, we compare our results with the calculations of other {\it{ab initio}} methods and discuss the role of central correlation in neutron stars. In Sect. 2, we give a basic theoretical formulation of the RCV method. In Sect. 3, numerical results are presented on the properties of neutron stars and are compared with other calculations. In Sect. 4, we give a summary of the present work.

\section{Relativistic central variational method}
We firstly show the Hamiltonian of the Bonn potential~\cite{machleidt89}, which is constructed based on the one-boson-exchange potential. It is defined as a sum of one-particle amplitudes of six bosons, consisting of $\pi$ and $\eta$ pseudovectors, $\sigma$ and $\delta$ scalars, and $\rho$ and $\omega$ vector mesons,
\beq
H&=&\sum^A_{i}T_i+\frac{1}{2}\sum^A_{i,j}V_{i,j}\nn
&=&\int d^3\mathbf{x}\overline \psi(\mathbf x)(-i\bm{\gamma}\cdot\bm{\nabla}+M_N)\psi(\mathbf x)\nn
&&+\frac{1}{2}\sum_{i=\sigma,\delta,\atop\eta,\pi,\omega,\rho}\int d^3\mathbf{x}'d^3\mathbf{x}\overline \psi(\mathbf x')\overline \psi(\mathbf x)\frac{\Gamma_i(1,2)}{m^2_i+\mathbf q^2}\psi(\mathbf x)\psi(\mathbf x'),
\eeq
where $\psi$ corresponds the nucleon field. $\mathbf q$ is the transfer momentum between two nucleons and $M_N,~m_i$ are the masses of nucleons and mesons, respectively. The $\Gamma_i$ matrices represent the vertex between nucleons and mesons, which are depicted below:
\beq
&&\Gamma_\sigma(1,2)=-g^2_\sigma,\nn
&&\Gamma_\delta(1,2)=-g^2_\delta\bm{\tau}_1\cdot\bm{\tau}_2,\nn
&&\Gamma_\eta(1,2)=-\left(\frac{f_\eta}{m_\eta}\right)^2(\rlap/q\gamma_5)_1(\rlap/q\gamma_5)_2,\nn
&&\Gamma_\pi(1,2)=-\left(\frac{f_\pi}{m_\pi}\right)^2(\rlap/q\gamma_5)_1(\rlap/q\gamma_5)_2\bm{\tau}_1\cdot\bm{\tau}_2,\nn
&&\Gamma_\omega(1,2)=g^2_\omega\gamma_\mu(1)\gamma^\mu(2),\nn
&&\Gamma^V_\rho(1,2)=g^2_\rho\gamma_\mu(1)\gamma^\mu(2)\bm{\tau}_1\cdot\bm{\tau}_2,\nn
&&\Gamma^T_\rho(1,2)=\left(\frac{f_\rho}{2M_N}\right)^2q_\nu\sigma^{\mu\nu}(1)q^\lambda\sigma_{\mu\lambda}(2)\bm{\tau}_1\cdot\bm{\tau}_2,\nn
&&\Gamma^{VT}_\rho(1,2)=i\left(\frac{g_\rho f_\rho}{M_N}\right)\gamma_\mu(2)\sigma^{\mu\nu}q_\nu(1)\bm{\tau}_1\cdot\bm{\tau}_2,
\eeq
where $\sigma_{\mu\nu}=\frac{i}{2}[\gamma_\mu,\gamma_\nu]$ is an antisymmetric tensor gamma matrix, and the tensor coupling part between $\omega$ meson and nucleon has been neglected because the value of $f_\omega/g_\omega$ is negligible. Meanwhile, a monopole form factor should be considered,
\beq
F_i(q^2)=\frac{\Lambda^2_i-m^2_i}{\Lambda^2_i+q^2},
\eeq
for each meson-nucleon vertex denoted by $i$. All coupling constants and cut-off momenta $\Lambda_i$ were determined by fitting the $NN$ scattering data.

In the RCV method, one introduces a central correlation function of the wave function of the RHF theory as the trial wave function of the nuclear matter system \cite{hu11}, to treat the strong short-range repulsion,
\beq
|\Psi\ket=F|\Phi\ket,
\eeq
where $|\Phi\ket$ is the RHF wave function, and the correlation factor $F$ is chosen to be a product of two-body correlation functions $f(r_{ij})$,
\beq
F=\prod^A_{i<j}f(r_{ij}).
\eeq
Here, $f(r_{ij})$ is Jastrow correlation function~\cite{jastrow55}. The total energy density with the correlation function is obtained as,
\beq
\mathcal{E}_c&=&\frac{E_c}{\Omega}=\frac{1}{\Omega}\bra\Phi|\widetilde{H}|\Phi\ket\nn
&=&\bra T\ket+\bra T_c\ket+\bra V\ket.
\eeq
where the explicit form of the correlated Hamiltonian $\widetilde{H}$ appears as
\beq
\widetilde{H}&=&\sum^A_iT_i+\frac{1}{2}\sum^A_{i,j}\widetilde {V}_{ij}\nn
&=&\sum^A_iT_i+\frac{1}{2}\sum^A_{i,j}\{f^\dag(r_{ij})[T_i+T_j+V_{ij}]f(r_{ij})-(T_i+T_j)\}.
\eeq
Furthermore, the kinetic energy density part is not only related with the original one-body kinetic operator $\bra T\ket$,
\beq
\bra T\ket=\frac{\lambda}{\pi^2}\int^{k_F}_0p^2dp[p\hat P+M_N\hat M],
\eeq
but also the two-body operator with the Jastrow correlation function, $\bra T_c\ket$,
\beq
\bra T_c\ket&=&C\rho_B\frac{\lambda}{\pi^2}\int^{k_F}_0p^2dp[p\hat P+M_N\hat M]\nn
&-&\frac{2\lambda}{(2\pi)^4}\int^{k_F}_0p^2dpp'^2dp'\{[p\hat P(p)+2M_N \hat M(p)]I(p,p')+p'\hat P(p)J(p',p)\},
\eeq
where the first and second terms correspond to the direct (Hartree) and exchange (Fock) contributions of $\bra T_c\ket$, respectively and $C=\int d^3\mathbf r[f^2(r)-1]$. The isospin degeneracy is $\lambda=2$ for symmetric nuclear matter and $\lambda=1$ for pure neutron matter.  $\hat P$ and $\hat M$ are defined as
\beq
\hat P(p)=\frac{p^*(p)}{E^*(p)},~~~~~~~~~~~~~~\hat M(p)=\frac{M^*_N(p)}{E^*(p)}.
\eeq
The potential energy density $\bra V\ket$ will be split into the direct term $\bra V_D\ket$ and exchange term $\bra V_E\ket$. The pseudo-vector mesons do not provide their contributions in the Hartree approximation. We have the result of $\bra V_D\ket$ as,
\beq
\bra V_D\ket&=&-\frac{\widetilde {F}_\sigma(0,0)}{2}(\rho^p_S+\rho^p_S)^2-\frac{\widetilde {F}_\delta(0,0)}{2}(\rho^p_S-\rho^n_S)^2\nn
&&+\frac{\widetilde {F}_\omega(0,0)}{2}(\rho^p_B+\rho^n_B)^2+\frac{\widetilde {F}_\rho(0,0)}{2}(\rho^p_B-\rho^n_B)^2,
\eeq
where $\rho_S$ and $\rho_B$ are the scalar and nucleon vector densities, respectively, given as,
\beq
\rho^{p(n)}_S&=&\frac{1}{\pi^2}\int^{k^{p(n)}_F}_0p^2dp\hat M(p),\nn
\rho^{p(n)}_B&=&\frac{1}{\pi^2}\int^{k^{p(n)}_F}_0p^2dp.
\eeq
The proton and neutron cases are distinguished by the superscript $p/n$. The exchange contribution of $\sigma$ meson $\bra V^\sigma_E\ket$ as an example is given by,
\beq
\bra V^\sigma_E\ket=\frac{\lambda}{2(2\pi)^4}\int^{k_F}_0pdpp'dp\left[A_\sigma(p,p')+\hat M(p)\hat M(p')B_\sigma(p,p')+\hat P(p)\hat P(p')C_\sigma(p,p')\right].
\eeq
Here, the moment dependent functions $A_i, B_i, C_i, I, J$ and $F_i$ are various angular integrals related with the meson-exchange potentials, which are listed in the appendix of Ref.~\cite{hu11}. The contributions of other mesons can be expressed in similar forms. Finally, the total energy density is written as,
\beq
\mathcal{E}_c=\bra T\ket+\bra T_c\ket+\bra V_D\ket+\sum_i\bra V^i_E\ket.
\eeq
Usually, several free parameters, $c_1, c_2,\dots,c_i$, will appear in the Jastrow function $f(r)$. In the present work, the correlation function is chosen as,
\beq
f(r)=1-(c_0+c_1r+c_2r^2+c_3r^3)e^{-c_4r}.
\eeq
where the exponential term makes $f(r)$ unity at large distance. A natural choice from the unitary property of the correlation function is a normalization constraint on $f(r_{ij})$,
\beq\label{nc}
\int d^3 r_{ij}[f^2(r_{ij})-1]=0,
\eeq
and should go to zero for small $r_{ij}$ because of the repulsive core of $NN$ interaction, which lead to $c_0=1$. Furthermore, we should also ensure it is a monotonously increasing property at short distance,
\beq\label{gcon}
f'(0)\geq 0.
\eeq
We can then calculate the binding energy of nuclear matter after determining the remaining parameters with the variational principle. The minimal value of the total energy should appear at $f'(0)=0$ and $f''(0)=0$ with the constraint (\ref{gcon}) to make the Jastrow correlation function increase at short distance. Therefore, we can obtain the relations between $c_1, c_2$, and $c_4$,
\beq
c_1=c_4,~~~~c_2=\frac{c^2_4}{2}.
\eeq
Now, there is only one parameter, $c_4$, in the actual calculation, because $c_3$ is fixed by the normalization condition of the Jastrow correlation function, Eq. (\ref{nc}). We would like to determine $c_4$ by the variational principle with the energy density,
\beq
\frac{\partial\mathcal{E}_c}{\partial c_4}=0.
\eeq
More detailed formulas of the RCV method can be found in Ref.~\cite{hu11}.

For neutron star matter, there are not only the nucleons, but also the leptons, like electrons and muons. All of them exist in the neutron star with the equilibrium conditions of the chemical potentials for the $\beta$ decay,
\beq\label{bteq}
\mu_n&=&\mu_p+\mu_e,\nn
\mu_\mu&=&\mu_e,
\eeq
where the chemical potentials $\mu_n,~\mu_p,~\mu_\mu$, and $\mu_e$ are determined by the relativistic energy-momentum relation at the Fermi momentum $p=k_F$,
\beq
\mu_i&=&\Sigma^i_0(k_F)+E^\ast_i(k_F),\nn
\mu_\lambda&=&\sqrt{k^2_{F,\lambda}+m^2_\lambda},
\eeq
where, $i=n,~p$ and $\lambda=e,~\mu$. $\Sigma^i_0(k_F)$ is the zero component of the self-energy of proton or neutron. Furthermore, the nucleon density conservation and charge neutrality are imposed in neutron star matter as,
\beq
&&\rho=\rho_{n}+\rho_{p},\nn
&&\rho_e+\rho_\mu=\rho_{p}.
\eeq
The pressure of the neutron star system can be obtained with thermodynamics relation, as
\beq
P(\rho)=\rho^2\frac{\partial}{\partial\rho}\frac{\varepsilon}{\rho}=\sum_{i=n,p,e,\mu}\rho_{i}\mu_i-\varepsilon.
\eeq

The stable configurations of a neutron star then can be obtained from the well known hydrostatic equilibrium equations, by Tolman, Oppenheimer and Volkoff~\cite{tolman39,oppenheimer39} for the pressure $P$ and the enclosed mass $m$,
\beq\label{tov}
\frac{dP(r)}{dr}&=&-\frac{Gm(r)\varepsilon(r)}{r^{2}}\frac{\Big[1+\frac{P(r)}{\varepsilon(r)}\Big]\Big[1+\frac{4\pi r^{3}P(r)}{m(r)}\Big]}
{1-\frac{2Gm(r)}{r}},\nn
\frac{dm(r)}{dr}&=&4\pi r^{2}\varepsilon(r),
\eeq
where, $P(r)$ is the pressure of neutron star at radius, $r$, and $m(r)$ is the total star mass inside a sphere of radius $r$. Once the EOS $P(\varepsilon)$ is specified, $\varepsilon$ being the total energy density ($G$ is the gravitational constant), for a chosen central value of the energy density, the numerical integration of Eq.(\ref{tov}) provides the mass-radius relation of neutron star.

\section{Results and discussions}
\subsection{The equations of state of neutron star matter}
\begin{figure}[!hbt]
	\centering
	\subfigure[]{\includegraphics[width=8cm]{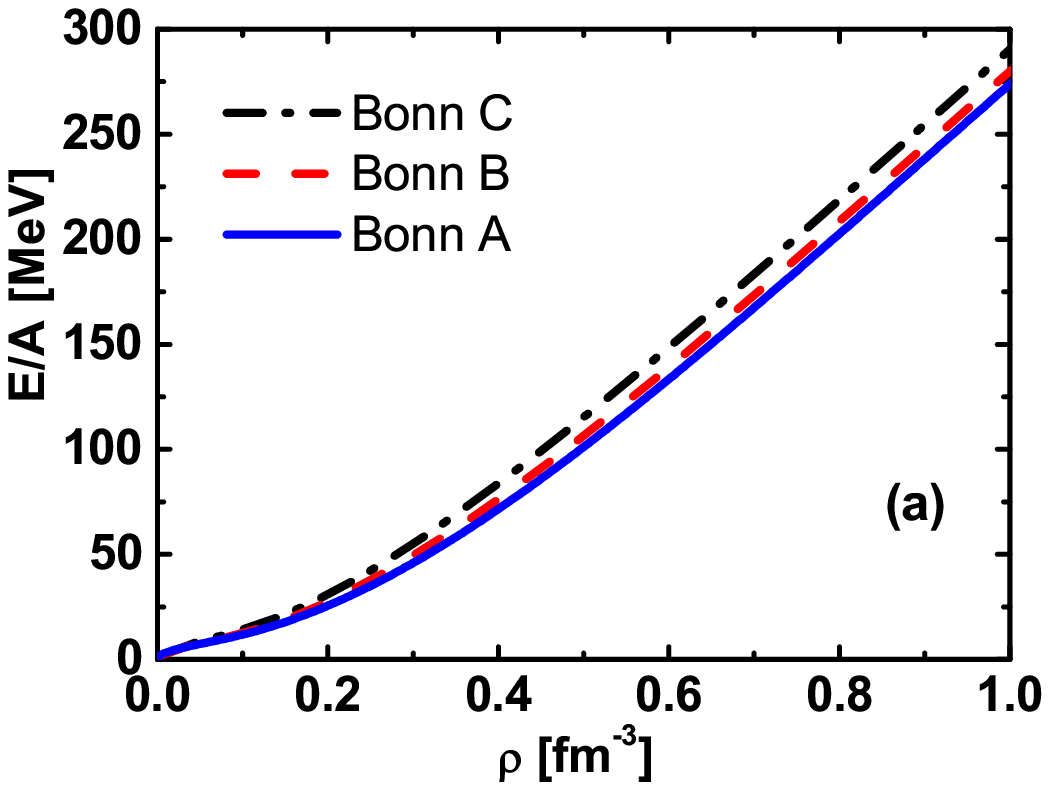}}
	\subfigure[]{\includegraphics[width=8cm]{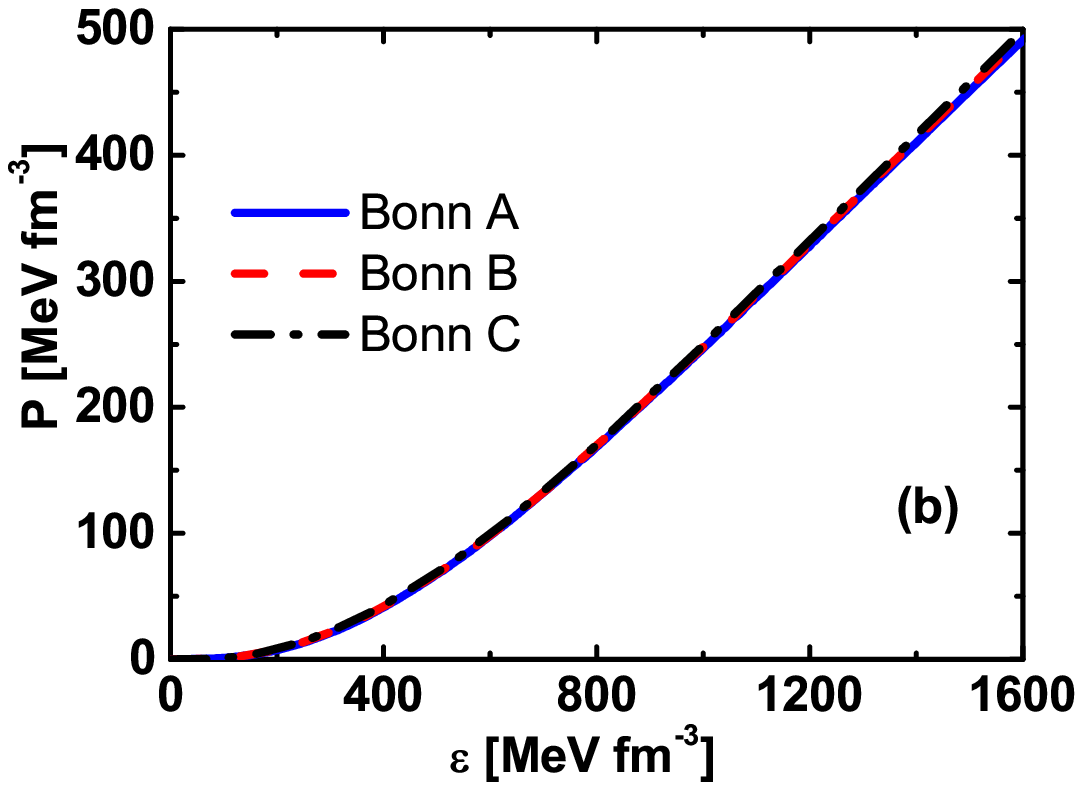}}
	\caption{The energy per nucleon as a function of density (a panel) and the pressure as a function of energy density (b panel) of neutron star matter with Bonn A, Bonn B, and Bonn C potentials.}
	\label{re}
\end{figure}

After solving the equilibrium conditions of chemical potentials for various particles in the neutron star matter, Eq.~(\ref{bteq}), the binding energies per nucleon as functions of density are shown in the left panel of Fig.~\ref{re} with three Bonn potentials, Bonn A, Bonn B, and Bonn C, which were fitted by the scattering data of $NN$ system. The binding energy of the Bonn A potential is the smallest of the three Bonn potentials, which is in accordance with the conclusion of RBHF theory for symmetric nuclear matter~\cite{brockmann90}. The tensor component of Bonn A among these three potentials is the weakest, whose $D$ state probability, $P_D$, is the smallest in deuteron. Furthermore, the difference of binding energies of these three potentials in neutron star matter is obviously less than that in the symmetric nuclear matter. The effect of the tensor force becomes weaker with increasing neutron fraction and does not play any role in pure neutron matter. Similarly, the pressure as a function of energy density for neutron star matter is required as input data, when we calculate the properties of neutron stars in the TOV equation. The pressure-energy relations with Bonn potentials are given in the right panel of Fig.~\ref{re}. Their behaviors are very similar to those of the binding energy and are almost identical with each other. It demonstrates that these EOSs will generate similar properties of the neutron stars.

\begin{figure}[!hbt]
	\centering
	\includegraphics[width=9cm]{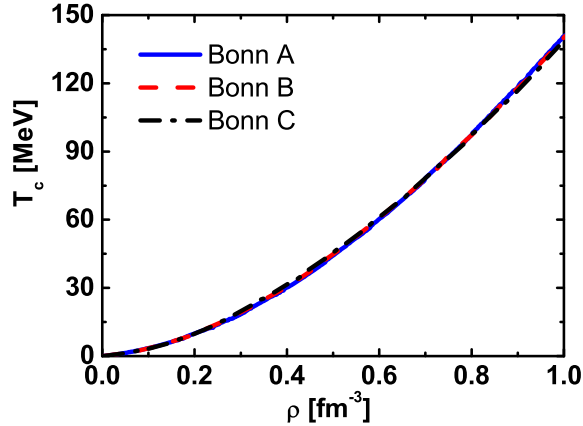}
	\caption{The central correlation on kinetic energy for neutron star matter with Bonn A, Bonn B, and Bonn C potentials.}
	\label{tc}
\end{figure}

The central correlation on kinetic energy per nucleon is presented for the neutron star matter in Fig.~\ref{tc}. The variational method can be achieved based on the competition between the correlation on kinetic energy and potential. The central correlation on kinetic energy provides a repulsive effect to prevent two-nucleon approach at high densities, whereas the one on potential gives an attractive contribution to remove the repulsion of realistic $NN$ interactions in short-range region. Finally, the correlations on kinetic and potential energies determine the minimum total energy and explicitly confirm the variational parameters in the Jastrow function. It can be found that the correlation on kinetic energy contributes almost half of the binding energy per nucleon for neutron star matter and plays a very essential role in the RCV method.

\begin{figure}[!hbt]
	\centering
	\includegraphics[width=9cm]{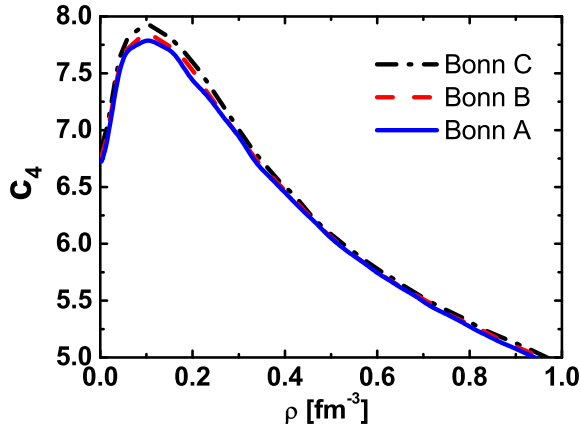}
	\caption{The variational parameter $c_4$ as a function of density for neutron star matter with Bonn A, Bonn B, and Bonn C potentials.}
	\label{c4}
\end{figure}

In the RCV method there is only one independent variational parameter, $c_4$~\cite{hu11}, which is shown in Fig.~\ref{c4} as a function of density with Bonn A, Bonn B, and Bonn C potentials. From this figure, we can see that the central correlation strength increases slowly at low density, reaches a maximum value around the normal nuclear saturation density, and starts to decrease with increasing density thereafter. It demonstrates that the central correlation will have to consider more variables to generate the saturation density and that it becomes weaker at high density, since the distance between two nucleons is already sufficiently compressed.
\subsection{The properties of neutron stars}
\begin{figure}[!hbt]
	\centering
	  \subfigure[]{\includegraphics[width=8cm]{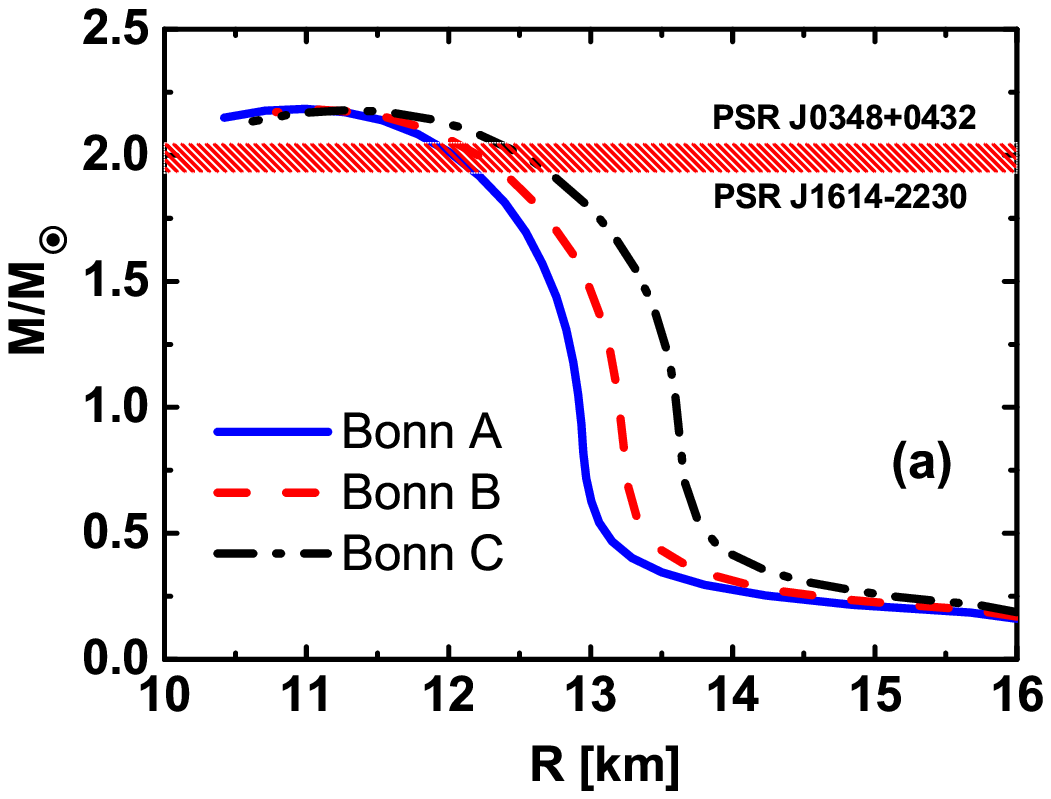}}
	  \subfigure[]{\includegraphics[width=8cm]{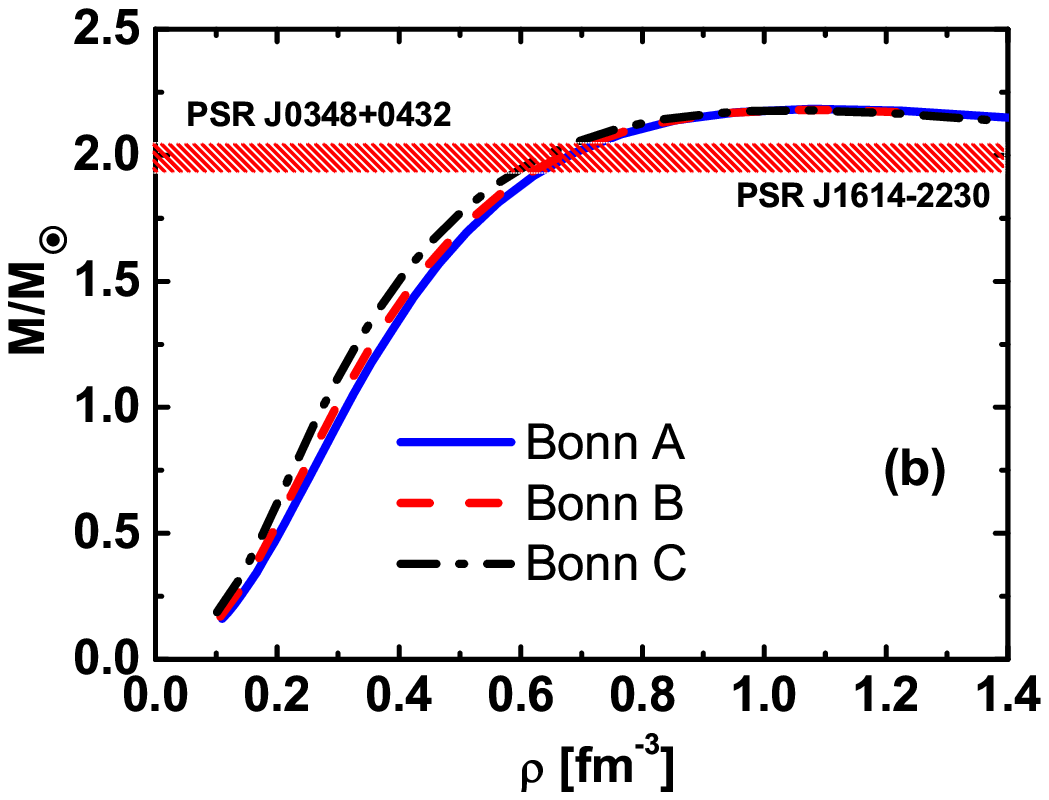}}
	\caption{The mass-radius relations (panel a) and mass-density relations (panel b) of neutron stars with Bonn A, Bonn B, and Bonn C potentials.}
	\label{rm}
\end{figure}

With the EOSs from the RCV method, we obtain the properties of neutron stars by solving the TOV equation with Bonn potentials. The mass-radius and mass-density relations of neutron stars within our present framework are illustrated in Fig.~\ref{rm}. In the left panel, we plot the mass-radius relations of neutron stars with Bonn A, Bonn B, and Bonn C potentials. The maximum masses and corresponding radii predicted are almost the same, around $2.18 M_\odot$ and $11$ km, respectively. These results are in good agreement with the previous {\it{ab initio}} calculations. The maximum masses of neutron stars are $2.2 M_\odot$ in non-relativistic variational method by Akmal {\it{et al.}}~\cite{akmal98}. Similarly, the  maximum masses and corresponding radii of neutron stars were given as $2.24 M_\odot$ and $10.8$ km in the RBHF theory with Bonn potentials~\cite{krastev06}. It indicates that our RCV method can economically describe the neutron star matter as well as both the non-relativistic full variational method and the RBHF theory. The maximum masses and the mass-radius relations of neutron stars with three Bonn potentials have obvious distinctions, since the neutron star matter includes not only neutrons but also protons. The tensor force should contribute to some effects at the $T=0$ channel here, especially at low density, which is shown clearly in the mass-radius curves at large $R$. In the right panel of Fig.~\ref{rm}, we plot the mass-density relations of neutron stars with Bonn potentials. The central densities at maximum neutron star mass are $1.078$ fm$^{-3}$ for all of three Bonn potentials, which are similar to the calculations of Akmal {\it{et al.}}~\cite{akmal98}, $1.14$ fm$^{-3}$ and RBHF theory, $1.003-1.013$ fm$^{-3}$~\cite{krastev06}. In the low density region, the masses of neutron stars display distinguishable behaviors. They are almost identical with increasing densities.

In Table~\ref{pro}, the properties of neutron stars within the present framework are listed for Bonn A, Bonn B, and Bonn C potentials. They are compared with the ones taken from the RBHF theory~\cite{krastev06}. No matter the maximum mass or the corresponding radii and central density, our results are only $3\%$ different from the ones calculated via the RBHF theory. Furthermore, the maximum  masses of neutron stars in our calculation satisfy the requirements of recent observation on massive neutron stars~\cite{demorest10,antoniadis13}, around $2M_\odot$. The corresponding radius is also located within the constraint region worked out by Hebeler {\it{et al.}}~\cite{hebeler10,hebeler13}.
\begin{table}[!hbt]
	\centering
	\begin{tabular}{c c c c c c c}
		\hline
		\hline
		Potentials&                                    &     RCV      &                        &                                    &RBHF            &                     \\
		          &~~$M_{\text{max}}$ ($M_{\odot}$) &     $R$(km)      &  $\rho_c$ (fm$^{-3}$) &~~$M_{\text{max}}$ ($M_{\odot}$)&     $R$(km)    &  $\rho_c$ (fm$^{-3}$)\\
		\hline
		
		Bonn A    &~~2.184                               &10.99         &1.078                  &~~2.240                               &10.74           &1.013 \\
		
		Bonn B    &~~2.181                               &11.08         &1.078                  &~~2.240                               &10.79           &1.008 \\
		
		Bonn C    &~~2.179                               &11.22         &1.078                  &~~2.238                               &10.83           &1.003 \\
		
		\hline
		\hline
	\end{tabular}
	\caption{The properties of neutron stars (maximum mass, corresponding radii and central density) within present framework and compared with the ones calculated by RBHF theory for Bonn A, Bonn B, and Bonn C potentials.}\label{pro}
\end{table}

\begin{figure}[!hbt]
	\centering
	\includegraphics[bb=0 90 290 240, width=0.7\textwidth]{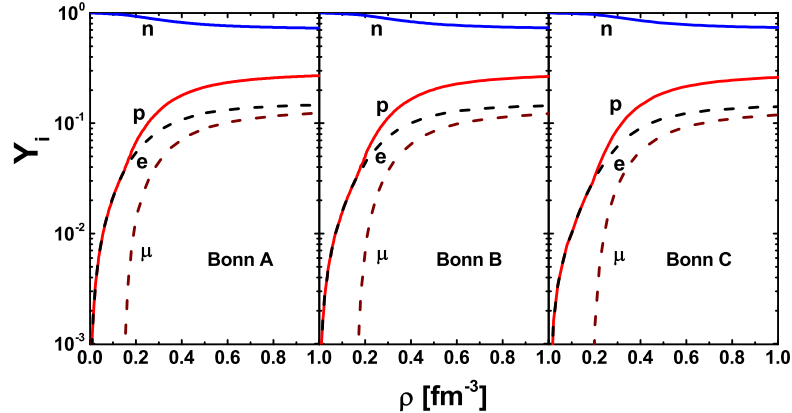}
	\caption{The particle fractions in neutron stars with Bonn A, Bonn B, and Bonn C potentials.}
	\label{ypabc}
\end{figure}

The fractions of various particles appearing in neutron stars, which are neutrons, protons, electrons, and muons are plotted in Fig.~\ref{ypabc} with the three Bonn potentials. At the beginning, the muon is absent for the $\beta$ equilibrium conditions. When the electron chemical potential is larger than the muon mass, the muon will appear in the neutron star matter at densities less than $0.2$ fm$^{-3}$, which is smaller than the density of muons in the RBHF theory. The earliest appearance of the muon in the Bonn A potential around the normal saturation density. At high density, the fraction of muons will approach that of electrons. The proton fraction in Bonn A has the largest magnitude compared to the other two Bonn potentials, which should be due to its smaller tensor component.

\begin{figure}[!hbt]
	\centering
	\includegraphics[width=9cm]{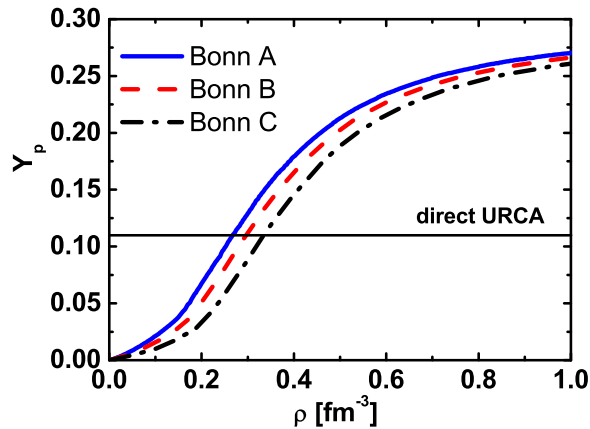}
	\caption{The proton fraction in neutron stars with Bonn A, Bonn B, and Bonn C potentials.}
	\label{urca}
\end{figure}

We give the proton fractions with Bonn potentials alone in Fig.~\ref{urca}. From this figure, the difference of proton fractions obtained from the three Bonn potentials is clearly revealed, which becomes very obviously in the intermediate density region and decreases in the high density region. The proton fraction in our calculation continues to increase with density and is in contrast to the case in RBHF theory, where it decreases at high density and the largest fraction is about $0.13$~\cite{krastev06}. In RBHF theory, to simplify the calculation, the leptons were treated with the non-relativistic approximation. At high densities, the relativistic effect becomes more important, where the Fermi momentum is very high and can be comparable with the light speed. Therefore, the direct URCA processes related with the cooling mechanism of neutron star could not occur easily with Bonn B and Bonn C potentials in RBHF theory, which should be satisfied by a proton fraction larger than approximately $1/9$. However, in our work, the direct URCA processes can be produced in all of the Bonn potentials and the densities appearing in direct URCA processes are located between $0.25-0.35$ fm$^{-3}$. These densities will lead the neutron star to cool very rapidly.
\subsection{The effective nucleon masses in neutron star}
\begin{figure}[!hbt]
	\centering
	\includegraphics[width=9cm]{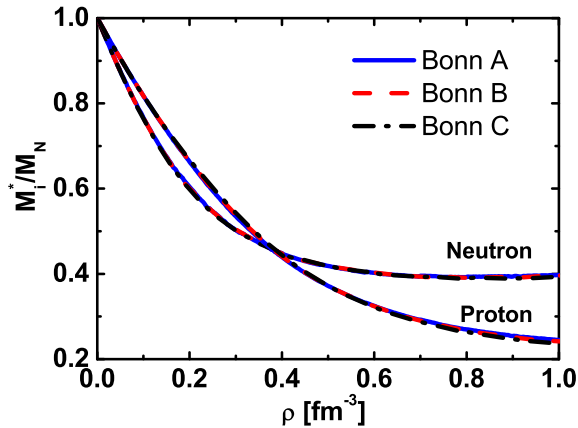}
	\caption{The effective proton and neutron masses in neutron stars with Bonn A, Bonn B, and Bonn C potential.}
	\label{emn}
\end{figure}

We also give the Dirac effective nucleon masses in neutron star matter in Fig.~\ref{emn}. For proton or neutron Dirac effective masses reflecting the nucleon media effect, their magnitudes with different potential are almost equal, which is very similar situation to the energy and pressure cases. If we compare the proton effective mass with the neutron effective mass, it can be found that the proton effective mass is larger than the neutron one at low density. When the density is $\rho>0.4$ fm$^{-3}$, the splitting of proton and neutron effective masses is reversed. This behavior is quite different from the effective mass in RHF theory, where the proton effective mass should be larger than the neutron one in all density regions~\cite{li16}. For RBHF theory, the splitting of proton and neutron effective masses for asymmetric nuclear matter is strongly dependent on the treatment of the $G-$matrix. Brockmann and Machleidt used the single-particle potential to extract the effective nucleon masses, where the neutron effective mass is larger than the proton one in neutron-rich matter~\cite{sammarruca10}; whereas Dalen {\it{et al.}} adopted the projection method to distinguish the spin components in the $G-$matrix and calculated the effective nucleon mass with these interaction components in the RHF model~\cite{dalen07,dalen10}. In this scheme, the proton effective mass is larger than the neutron one.

\begin{figure}[!hbt]
	\centering
	\includegraphics[width=9cm]{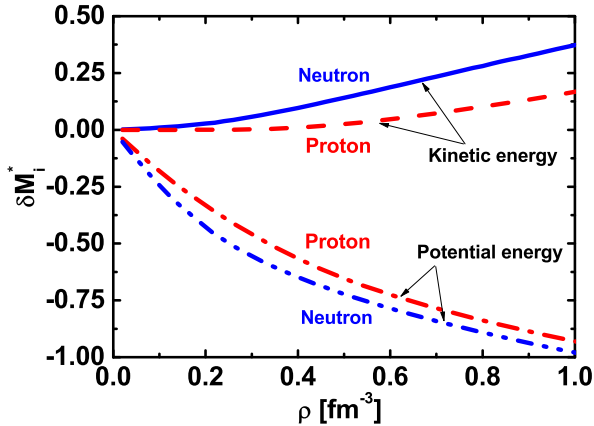}
	\caption{The kinetic and potential contributions on the nucleon effective masses in neutron star with Bonn A potential.}
	\label{emc}
\end{figure}

To discuss the splitting of proton-neutron effective masses in our framework, we show the kinetic and potential contributions of the nucleon effective mass in a neutron star with Bonn A potential in Fig.~\ref{emc}.  Without the central correlation function, the kinetic energy is a one-body operator, which does not provide any contribution to the nucleon effective mass. Once the central correlation function is included, the central correlation on kinetic energy becomes a two-body operator and contributes to the nucleon effective mass. From Fig.~\ref{emc}, we can find that the potential generates the negative contribution to the effective nucleon mass. In neutron-rich matter, the neutron effective mass will obtain more negative components compared to the proton one. Therefore, in the RHF model, the proton effective mass is larger than the neutron one in neutron-rich matter. Meanwhile, the central correlation on kinetic energy has the positive contributions to effective nucleon mass. Furthermore, its effect on neutrons is much higher than on protons at high densities. With the competition between the potential and kinetic energy, the proton effective mass is larger than the neutron one at low density and with increasing density, the neutron effective mass is larger than the proton one. Therefore, the central correlation on kinetic energy plays a very important role in the splitting of proton-neutron effective masses. For Dalen {\it{et al.}}, they obtained the effective nucleon mass from the potential part; whereas Brockmann and Machleidt considered the splitting of effective masses from the single particle potential, which is related to kinetic energy. This may be the reason why there are opposite conclusions between the two groups, in RBHF theory on effective nucleon mass.

\section{Conclusion}
The RCV method based on the framework of RHF theory was applied to study the properties of neutron stars with one-boson-exchange potentials, i.e., Bonn potentials, which were determined by fitting the nucleon-nucleon ($NN$) scattering data. In neutron-rich matter, the tensor force has a very small effect of its isospin feature. Therefore, it is essential to take the central correlation on the strong repulsion of $NN$ interaction at short distances for the description of neutron-rich matter. The equation of state (EOS) of pure neutron matter obtained by a novel {\it{ab initio}} calculation of RBHF theory was completely reproduced by the present RCV method.

The EOSs of neutron star matter in $\beta$ equilibrium with nucleons and leptons, were self-consistently solved in the RCV method with Bonn A, Bonn B, and Bonn C potentials. Their behaviors were almost identical for the weak effect of tensor force in neutron-rich matter, since the difference among three Bonn potentials only appeared in their tensor components. The relativistic central correlation on kinetic energy played a very important role in the process of minimizing the total binding energy, with the variational principle, and gave half of the contribution to total binding energy. Furthermore, we found that the strength of the central correlation function was the strongest at the saturation density through the variational parameters, which correspond to the saturation mechanism of symmetric nuclear matter.

The properties of neutron stars were studied with the EOSs of neutron star matter by solving the TOV equation. The maximum neutron star masses and corresponding radii were around $2.18 M_\odot$ and $11$ km, respectively, using the RCV method with Bonn potentials. The central densities of neutron stars were about $1.078$ fm$^{-3}$. These results are in good agreement with the calculations from RBHF theory and the non-relativistic variational method. It demonstrated that the RCV method can describe neutron-rich matter reasonably and economically, compared with the conventional {\it{ab initio}} calculation.

The proton fractions in neutron star matter with the three Bonn potentials showed some differences. The proton fraction in Bonn A potential was the largest and Bonn C the smallest. The direct URCA processes would be generated in all of these potentials above the densities, $0.25-0.35$ fm$^{-3}$ with a proton fraction larger than approximately $1/9$. Furthermore, the splitting of proton-neutron effective masses was reversed with increasing density of neutron star matter. In the low density region, the proton effective mass was larger than the neutron one; whereas these behaviors are opposite at high densities, since the central correlation on kinetic energy played a more important role at high densities compared with the correlation on potential. This may explain the conflicting predictions about the splitting of proton-neutron masses in the two different treatments in RBHF theory.

Although, we can describe the properties of neutron stars very well, which are comparable to the results obtained by the other {\it{ab initio}} methods, it is necessary to take the tensor correlation into account to study symmetric nuclear matter and its saturation properties to reproduce the empirical data in future.

\section*{Acknowledgments}
This work was supported in part by the National Natural Science Foundation of China (Grant No. 11375089 and Grant No. 11405090).


\begin{thebibliography}{99}
\bibitem{glendenning97}	N. K. Glendenning, {\it{Compact Stars, Nuclear Physics, Particle Physics, and General Relativity}} (Springer-Verlag, New York, 1997).
\bibitem{weber99} F. Weber, {\it{Pulsars as Astrophysical Laboratories for Nuclear and Particle Physics}} (IOP, Bristol, 1999).
\bibitem{prakash97} M. Prakash, I. Bombaci, M. Prakash, P. J. Ellis, J. M. Lattimer, and R. Knorren, Phys. Rep. {\bf 280}, 1 (1997).
\bibitem{heiselberg00} H. Heiselberg and M. Hjorth-Jensen, Phys. Rep. {\bf 328}, 237 (2000).
\bibitem{oppenheimer39} J. Oppenheimer and G. Volkoff, Phys. Rev. {\bf 55}, 374 (1939).
\bibitem{tolman39} R. C. Tolman, Phys. Rev. {\bf 55}, 364 (1939).
\bibitem{dutra12} M. Dutra, O. Lourenco, J. S. S\'a Martins, A. Delfino, J. R. Stone, and P. D. Stevenson, Phys. Rev. C {\bf 85}, 035201 (2012).
\bibitem{dutra14} M. Dutra, O. Lourenco, S. S. Avancini, B. V. Carlson, A. Delfino, D. P. Menezes, C. Provid\^{e}ncia, S. Typel, and J. R. Stone, Phys. Rev. C {\bf 90}, 055203 (2014).
\bibitem{vautherin72} D. Vautherin and D. M. Brink, Phys. Rev. C {\bf 5}, 626 (1972).
\bibitem{bender03} M. Bender, P. H. Heenen, and P. G. Reinhard, Rev. Mod. Phys. {\bf 75}, 121 (2003).
\bibitem{stone07} J. R. Stone and P. G. Reinhard, Prog. Part. Nucl. Phys. {\bf 58}, 587 (2007).
\bibitem{serot86} B. D. Serot and J. D. Walecka,  Adv. Nuc. Phys. {\bf 16}, 1 (1986).
\bibitem{ring96} P. Ring, Prog. Part. Nucl. Phys. {\bf 37}, 193 (1996).
\bibitem{meng06} J. Meng, H. Toki, S. G. Zhou, S. Q. Zhang, W. H. Long, and L. S. Geng, Prog. Part. Nucl. Phys. {\bf 57}, 470 (2006).
\bibitem{demorest10} P. B. Demorest, T. Pennucci, S. M. Ransom, M. S. E. Roberts, and J. W. T. Hessels, Nature (London) \textbf{467}, 1081 (2010).
\bibitem{antoniadis13} J. Antoniadis, P. C. C. Freire, N. Wex, T. M. Tauris, R. S. Lynch \textit{et al.}, Science \textbf{340}, 6131 (2013).
\bibitem{machleidt89} R. Machleidt, Adv. Nucl. Phys. {\bf19}, 189 (1989).
\bibitem{epelbaum09} E. Epelbaum, H.-W. Hammer, and Ulf-G. Mei{\ss}ner, Rev. Mod. Phys. {\bf81}, 1773, (2009).
\bibitem{machleidt11} R. Machleidt and D. R. Entem, Phys. Rep. {\bf503}, 1 (2011).
\bibitem{jastrow55} R. Jastrow,  Phys. Rev. {\bf 98}, 1479 (1955).
\bibitem{toki80} H. Toki, Z. Physik  A {\bf 294}, 173 (1980).
\bibitem{oka81} M. Oka and K. Yazaki, Prog. Theor. Phys. {\bf 66}, 556 (1981).
\bibitem{pandharipande79}V. R. Pandharipande and R. B. Wiringa, Rev. Mod. Phys. {\bf 51}, 821 (1979).
\bibitem{wiringa95} R. B. Wiringa, V. G. J. Stoks, and R. Schiavilla,  Phys. Rev. C {\bf 51}, 38 (1995).
\bibitem{akmal98} A. Akmal, V. R. Pandharipande, and D. G. Ravenhall,  Phys. Rev. C {\bf 58}, 1804 (1998).
\bibitem{brockmann90} R. Brockmann and R. Machleidt,  Phys. Rev. C {\bf 42}, 1965 (1990).
\bibitem{krastev06} P. G. Krastev and F. Sammarruca,  Phys. Rev. C {\bf 74}, 025808 (2006).
\bibitem{gandolfi09} S. Gandolfi, A. Yu. Illarionov, K. E. Schmidt, F. Pederiva, and S. Fantoni, Phys. Rev. C {\bf79}, 054005 (2009).
\bibitem{hebeler10} K. Hebeler, J. M. Lattimer, C. J. Pethick, and A. Schwenk, Phys. Rev. Lett. {\bf 105}, 161102 (2010).
\bibitem{hebeler13} K. Hebeler, J. M. Lattimer, C. J. Pethick, and A. Schwenk, Astrophys. J. {\bf 773}, 11 (2013).
\bibitem{hu10} J. Hu, H. Toki, W. Wen, and H. Shen,  Phys. Lett. B {\bf 687}, 271 (2010).
\bibitem{wang12} Y. Wang, J. Hu, H. Toki, and H. Shen, Prog. Theo. Phys. {\bf 127}, 739 (2012).
\bibitem{hu13} J. Hu, H. Toki, and Y. Ogawa, Prog. Theor. Exp. Phys. {\bf 103D02}, (2013).
\bibitem{hu11} J. Hu, H. Toki, and H. Shen, J. Phys. G. {\bf 38}, 085105 (2011).
\bibitem{panda05} P. K. Panda, D. P. Menezes, C. Provid\^{e}ncia, and J. da Provid\^{e}ncia, Phys. Rev. C {\bf71}, 015801 (2005).
\bibitem{panda06} P. K. Panda, J. da Provid\^{e}ncia, and C. Provid\^{e}ncia, Phys. Rev. C {\bf73}, 035805 (2006).
\bibitem{panda07} P. K. Panda, C. Provid\^{e}ncia, and J. da Provid\^{e}ncia, Phys. Rev. C {\bf75}, 065806 (2007).
\bibitem{bouyssy87} A. Bouyssy, J. F. Mathiot, N. Van Giai, and S. Marcos,  Phys. Rev. C {\bf 36}, 380 (1987).
\bibitem{li16} A. Li, J. N. Hu, X. L. Shang, and W. Zuo, Phys. Rev. C {\bf 93}, 015803 (2016).
\bibitem{sammarruca10} F. Sammarruca, Int. J. Mod. Phys. E {\bf 19}, 1259 (2010).
\bibitem{dalen07} E. N. E. Dalen, C. Fuchs, and A. Faessler,  Eur. Phys. J.  A {\bf 31}, 29 (2007).
\bibitem{dalen10} E. N. E. Dalen and H. Muether, Int. J. Mod. Phys. E {\bf 19}, 2077 (2010).







\end{thebibliography}
\end{document}